\documentclass[10pt,doublecolumn,twoside]{IEEEtran}

\usepackage{etoolbox}
\newtoggle{doublecolumn}
\newtoggle{calculateFigures}

\toggletrue{doublecolumn}
\togglefalse{calculateFigures}
\usepackage{etex}

\usepackage[english]{babel}
\usepackage{lipsum}
\usepackage{amsbsy,amsmath,amsfonts,amssymb,amsthm}
\usepackage{mathtools}
\usepackage{textcomp} 
\usepackage{relsize}

\usepackage{bm,cite,cases,pstricks,times,url,verbatim} 
\usepackage[noend]{algpseudocode}
\usepackage{booktabs}
\usepackage{tabularx,array,dcolumn,multirow}

\usepackage{yfonts}

\usepackage{dutchcal} 

\usepackage{soul} 

\usepackage{blkarray,bigdelim} 

\allowdisplaybreaks[4]

\usepackage{centernot} 

\newcommand{\subparagraph}{}

\iftoggle{doublecolumn}{
    \newtheorem{thm}{Theorem}
    \newtheorem{fact}{Fact}
    \newtheorem{lemma}{Lemma}
    \newtheorem{definition}{Definition}
    \newtheorem{conj}{Conjecture}
    \newtheorem{propos}{Proposition}
    \newtheorem{corol}{Corollary}
    \newtheorem{ass}{Assumption}
    \newtheorem{example}{Example}
    \newtheorem{remark}{Remark}
    \newtheorem{note}{Note}
    \newtheorem{obs}{Observation}
}{

    \newtheoremstyle{exampstyle}
      {0} 
      {0} 
      {\itshape} 
      {} 
      {\bfseries} 
      {.} 
      {.5em} 
      {} 

    \theoremstyle{exampstyle} 
    \theoremstyle{exampstyle} 
    \theoremstyle{exampstyle} 
    \theoremstyle{exampstyle} 
    \theoremstyle{exampstyle} 
    \theoremstyle{exampstyle} \newtheorem{propos}{Proposition}
    \theoremstyle{exampstyle} 
    \theoremstyle{exampstyle} 
    \theoremstyle{exampstyle} 
    \theoremstyle{exampstyle} 
    \theoremstyle{exampstyle} 
    \theoremstyle{exampstyle} 
}

\newcommand{\argmax}[1]{\underset{#1}{\operatorname{arg}\,\operatorname{max}}\;}

\makeatletter
\newcommand{\pushright}[1]{\ifmeasuring@#1\else\omit\hfill$\displaystyle#1$\fi\ignorespaces}
\newcommand{\pushleft}[1]{\ifmeasuring@#1\else\omit$\displaystyle#1$\hfill\fi\ignorespaces}

\begingroup
\catcode`\#=11
\gdef\noautorotate{-dAutoRotatePages#/None}
\endgroup

\usepackage{graphicx,xcolor,float,dblfloatfix}
\usepackage{psfrag}

\usepackage{caption}
\usepackage{subcaption}

\newcommand{\subalign}[1]{%
  \vcenter{%
    \Let@ \restore@math@cr \default@tag
    \baselineskip\fontdimen10 \scriptfont\tw@
    \advance\baselineskip\fontdimen12 \scriptfont\tw@
    \lineskip\thr@@\fontdimen8 \scriptfont\thr@@
    \lineskiplimit\lineskip
    \ialign{\hfil$\m@th\textstyle##$&$\m@th\textstyle{}##$\crcr
      #1\crcr
    }%
  }
}

\iftoggle{calculateFigures}{
    \usepackage[crop=off,runs=2,pspdf=\noautorotate]{auto-pst-pdf}
}{
    \usepackage[off]{auto-pst-pdf}
}

\usepackage{hyperref}

\graphicspath{%
{./Figures/}
}

\usepackage{caption}
\begin{document}

\author{\IEEEauthorblockN{Alessandro~Biason and~Michele~Zorzi}\\
\IEEEauthorblockA{\{biasonal,zorzi\}@dei.unipd.it\\
Department of Information Engineering, University of Padova - via Gradenigo
6b, 35131 Padova, Italy
}%
}

\title{On the Effects of Battery Imperfections in an Energy Harvesting Device}

\maketitle
\thispagestyle{empty}
\pagestyle{empty}

\begin{abstract}
Energy Harvesting allows the devices in a Wireless Sensor Network to recharge their batteries through environmental energy sources. While in the literature the main focus is on devices with ideal batteries, in reality several inefficiencies have to be considered to correctly design the operating regimes of an Energy Harvesting Device (EHD). In this work we describe how the throughput optimization problem changes under \emph{real battery} constraints in an EHD. In particular, we consider imperfect knowledge of the state of charge of the battery and storage inefficiencies, \emph{i.e.}, part of the harvested energy is wasted in the battery recharging process. We formulate the problem as a Markov Decision Process, basing our model on some realistic observations about transmission, consumption and harvesting power. We find the performance upper bound with a real battery and numerically discuss the novelty introduced by the real battery effects. We show that using the \emph{old} policies obtained without considering the real battery effects is strongly sub-optimal and may even result in zero throughput.
\end{abstract}

\section{Introduction}

Energy Harvesting (EH), thanks to its ability to prolong the network lifetime, has been widely studied in the last years and a lot of research is still being conducted on this topic. If equipped with harvesting capabilities, the nodes in a Wireless Sensor Network (WSN) can recharge their batteries with renewable energy sources. Since one of the main goals in a WSN is to keep the network operational for a very long time (even decades), energy harvesting is a promising technique to achieve such target. While, ideally, with EH the network may operate for an unlimited amount of time, in practice the network lifetime is constrained by hardware failures, \emph{e.g.}, the batteries have a limited number of charging/discharging cycles~\cite{Michelusi2013c}. Also, generally, the harvested power is low with respect to the device energy consumption. Because of this,  energy has to be carefully managed in order to optimally exploit the available resources.

Several energy sources have been considered in the past. The most studied one is sunlight because it provides high power harvesting rate and the technology is mature and efficient~\cite{Raghunathan2005}. Examples of other techniques~\cite{Chalasani2008} are piezoelectric, vibrations, wind, ambient RadioFrequency (RF), acoustic noise, etc. Different optimization problems related to EH can be found in the literature so far. Generally, the main focus is on optimizing some network parameters, \emph{e.g.}, throughput, delay, packet drop rate, etc. A lot of research focused on \emph{offline} optimization, where it was assumed that the Energy Harvesting Device (EHD) knows everything about the environment (future energy arrivals, past history, channel status, etc.)~\cite{Yang2012}. Some relevant papers were produced by Sharma \emph{et al.}~\cite{Sharma2010}, where the authors found throughput and mean delay optimal policies when the considered device has an infinite data queue. Also, Ozel \emph{et al.} in~\cite{Ozel2012b} set up the offline throughput optimization problem and presented schemes to achieve the AWGN capacity. Tutuncuoglu \emph{et al.} studied offline policies for more than one device~\cite{Tutuncuoglu2012}. The authors also presented several works on the imperfect storage capabilities of energy harvesting devices~\cite{Tutuncuoglu2015}.

Another class of problems regards online optimization. In this context, only a statistical knowledge of the environment is required. Generally, a Markov approach is used to model and solve the problem~\cite{Lei2009}. \cite{Michelusi2013} considered the case of a correlated energy generation process. \cite{Li2011} studied the use of relays in an energy harvesting communication system. The interactions among multiple devices were analyzed in~\cite{Biason2014,Biason2015e,Michelusi2015}.

Despite the many works that can be found in this area, only few of them consider a non-ideal device. Indeed, several aspects are frequently neglected when designing the optimal policy, \emph{e.g.}, battery degradation~\cite{Michelusi2013c}, energy leakage~\cite{Devillers2012}, imperfect knowledge of the State Of Charge (SoC)~\cite{Michelusi2014}, storage losses~\cite{Biason2015c,Gorlatova2013,Tutuncuoglu2015} or circuitry costs~\cite{Ni2015}. In reality, the EHD battery is affected by all these aspects and, in general, the optimal policy obtained without considering them is not necessarily optimal. In this work we redefine the optimization problem when the device 1) has only a limited knowledge of its SoC and 2) the battery has storage inefficiencies (possibly dependent upon the current SoC). Even if in the literature these problems were partially studied separately, the combination of the two represents a more realistic case. Moreover, we find that the existing policies for the two separate scenarios are not applicable to the combined case because they would provide very poor performance. To model the system, we use an online approach and focus on the throughput optimization problem. In the definition of the system model we take into account some realistic and practical considerations (for example, using the energy consumption of a real device). We present two heuristic sub-optimal policies and discuss how they can be applied to our case. In our numerical evaluation we remark the importance of computing a new optimal policy, explicitly designed for the imperfect SoC case with a real battery.

The paper is organized as follows. Section~\ref{sec:system_model} defines the system model. The optimization model, the performance upper bound and the sub-optimal policies are discussed in Section~\ref{sec:optimization}. Section~\ref{sec:numerical_evaluation} presents our numerical results. Section~\ref{sec:conclusions} concludes the paper.

\section{System Model}\label{sec:system_model}
In this paper, we focus on an Energy Harvesting Device (EHD) with energy losses in the battery storage process and imperfect knowledge of its State of Charge (SoC).

Time is divided in frames where frame $k$ corresponds to the time interval $\big[Tk,T(k+1)\big)$. At the beginning of every frame, EHD decides whether to stay idle or become active. In the idle phase, the energy consumption is assumed negligible. In the active phase, a stream of bits is sent to an external receiver during the first $\delta < T$ seconds of the frame. We assume that $\delta$ is much lower than $T$. This is a realistic assumption that will be discussed in more detail in Section~\ref{sec:energy_consumption}.\footnote{It is also possible to extend the model to the case $\delta \approx T$ by redefining the function $y_T$ of Equation~\eqref{eq:y_t}.}

\subsection{State of Charge}
The device has the capability of gathering energy from the environment and store it in a finite battery. We model the several energy quantities in the system as discrete \emph{energy quanta}.\footnote{If the quantization granularity is sufficient, the discrete model can be considered as a good approximation of the continuous case.} The device has a battery of finite size $e_{\rm max} \in \mathbb{N}\backslash \{0\}$ energy quanta. In every time frame, the energy stored in the battery is $E_k \in \mathcal{E} \triangleq \{0,\ldots,e_{\rm max}\}$. As in~\cite{Michelusi2014}, we assume that, in the general case, EHD cannot read all the values of $\mathcal{E}$, but can only observe the partition $\tilde{\mathcal{T}}$ (with $N$ subsets). For example, the node may only know if the energy level is low or high ($N = 2$). In this case, we have $\tilde{\mathcal{T}} = \{{\rm LOW},{\rm HIGH}\}$, with ${\rm LOW} \triangleq \{0,\ldots,\lfloor e_{\rm max}/2 \rfloor\}$ and ${\rm HIGH} \triangleq \{\lfloor e_{\rm max}/2 \rfloor+1,\ldots,e_{\rm max}\}$. When the battery level is $e \in \mathcal{E}$, the node can observe only $\tilde{t} \in \tilde{\mathcal{T}}$, where $\tilde{t}$ is given by $\tilde{t} = \psi(e)$. In the previous example, $\psi(e) = {\rm LOW},\ \forall e \leq \lfloor e_{\rm max}/2\rfloor$ and $\psi(e) = {\rm HIGH},\ \forall e \geq \lfloor e_{\rm max}/2\rfloor +1$.\footnote{Note that the perfect SoC case is obtained when the partition $\tilde{\mathcal{T}}$ has $e_{\rm max}+1$ subsets (one for every energy state) and the function $\psi(e) = \{e\}$.} 

In our work we study the case where $\tilde{\mathcal{T}}$ is a partition of $\mathcal{E}$, but future work includes the extension to the case where different subsets of $\tilde{\mathcal{T}}$ are partially overlapped. This may be useful to study the cases where only a noisy observation of $e$ is available.

Note that it is meaningful to consider the imperfect SoC case because, in reality, the battery level cannot be known with absolute precision but can only be approximated (this strongly depends upon the considered technology). Moreover, we will find that, even restricting the study to the imperfect knowledge case, high performance can be achieved.

\subsection{Harvested Energy}
The energy arrivals are modeled as a random i.i.d. process (the case with memory could be studied with an approach similar to~\cite{Michelusi2014} but, because of the curse-of-dimensionality, can be solved only with sub-optimal techniques). Thus, in frame $k$, $B_k$ energy quanta arrive at the device, according to some energy arrival statistics with probability mass function (pmf) $p_B(b)$, mean $\bar{b}$ and maximum arrivals $b_{\rm max}$. In an ideal battery (no losses), all the $B_k$ energy quanta can be stored in the battery. In this work, we consider a \emph{real battery} with losses in the energy storage process. Different energy loss models exist, \emph{e.g.}, we can assume that only a fixed fraction $\eta$ of the incoming power can be stored~\cite{Tutuncuoglu2015}, or, more generally, that $\eta$ depends upon the current state of charge of the battery. This is a realistic assumption, \emph{e.g.}, in a capacitor~\cite{Gorlatova2013}. In this case, when the state of charge $E_k$ is low or high, then only a small fraction of the incoming energy can be stored, whereas, if $E_k \approx e_{\rm max}/2$, then almost all the energy can be successfully stored. An example, that was proposed in \cite{Gorlatova2013} as an approximation for the storage losses in a capacitor and will be used in our numerical evaluations (however, our results are general and do not depend upon the particular structure of $\eta(e)$), is~
\begin{align}
    \eta(e) &= 1-\frac{(e-e_{\rm max}/2)^2}{\beta_{\rm n.l.} (e_{\rm max}/2)^2}, \label{eq:s_x_y} 
\end{align}

\begin{figure}[t]
  \centering
  \includegraphics[trim = 0mm 0mm 0mm 3mm,  clip, width=1\columnwidth]{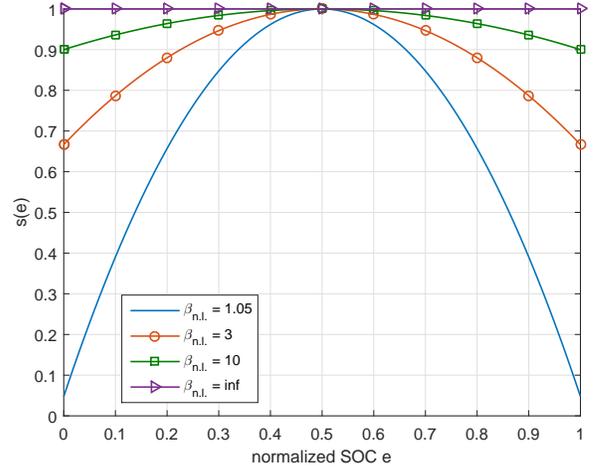}
  \caption{Function $\eta(e)$ (Equation~\eqref{eq:s_x_y}) as a function of the normalized SoC when $e_{\rm max} = 100$ (however, the structure of $\eta(e)$ is only weakly dependent on $e_{\rm max}$).}
  \label{fig:plot_s_y}
\end{figure}

\noindent where $\beta_{\rm n.l.} > 1$ is a constant which strictly depends upon the used technology. An example, that we will use as a baseline in this work, is $\beta_{\rm n.l.} = 1.05$ (see~\cite{Gorlatova2013}). In \figurename~\ref{fig:plot_s_y} we plot $\eta(e)$ as a function of the state of charge $e \in \mathcal{E}$, normalized with respect to $e_{\rm max}$. Note that, as $\beta_{\rm n.l.}$ increases, the storage losses decrease, until the asymptotic situation $\beta_{\rm n.l.} \rightarrow \infty$ where we obtain an ideal battery with no losses. 

Assume that a harvesting power $B_k/T$ is received (constant over a frame). In this case, only a power $\eta(e)B_k/T$ can be converted and stored in the battery. Note that, at $e = e_{\rm max}/2$ there are no energy losses, \emph{i.e.}, $\eta(e) = 1$, whereas when $e \in \{0,e_{\rm max}\}$, the losses are highest, \emph{i.e.}, $\eta(e) = 1-1/\beta_{\rm n.l.}$.

We now describe how to evaluate the stored energy evolution during a single frame. Equation~\eqref{eq:s_x_y} is an instantaneous expression that has to be applied only to \emph{powers}. Assume that during a frame the harvested power is constant and equal to $B_k/T$ and the battery is charged at $E_k$. After $t \leq T$ seconds, the battery status will ideally be~
\begin{align}
    y_t = E_k + \frac{1}{T}\int_0^t B_k \eta(y_\tau)\mbox{d}\tau. \label{eq:y_t}
\end{align}

At $t = T$ we obtain the battery level after an entire frame, \emph{i.e.}, $E_{k+1} = y_T(B_k,E_k)$. Also, since the battery energy levels are discretized, we approximate $y_T(B_k,E_k)$ with a $Round(\cdot)$ operation. In Equation~\eqref{eq:batt_update} we will present the full expression of the battery evolution by also considering other phenomena (overflow and transmission energy).

Note that the procedure to evaluate $y_T(B_k,E_k)$ does not depend upon the particular structure of $\eta(e)$. The only basic hypothesis that has to be made is that $Round(y_T(b_{\rm max},e)) \geq 1$ for every $e$ (otherwise, there may exist a state where recharging the battery would not be possible). However, it is always possible to satisfy this hypothesis by redefining the notion of energy quantum or the frame length.

\subsection{Energy Consumption}\label{sec:energy_consumption}
In frame $k$, a power $P_k \in \mathcal{P}$ (\emph{e.g.}, $\mathcal{P} \triangleq \{0,\rho_{\rm min},\ldots,\rho_{\rm max}\}$) is used for transmitting a new stream of bits (we assume that in every frame there is always enough data to be transmitted). Transmitting with a power $\rho \geq 0$ provides a reward $r(\rho)$, where $r(\cdot)$ is the \emph{instantaneous reward function}, positive, concave, increasing in $\rho$ and with $r(0) = 0$. A typical example of $r(\rho)$, that we also use as a baseline in our work, is $r(\rho) = \ln(1+\Lambda \rho)$, where $\Lambda$ represents an SNR scaling factor~\cite{Sharma2010,Ozel2012b,Michelusi2013,Blasco2013}. In this case, $r(\cdot)$ represents the transmission rate.

In slot $k$, the device spends $D_k$ energy quanta in order to perform the transmission. In order to consider the circuitry costs, in general $D_k > P_k$ (since the transmission duration $\delta$ is fixed, we use the terms ``power'' and ``energy'' interchangeably). An example of power consumption and corresponding transmission power is given in Table~\ref{tab:MCU}.

\vspace{0.3cm}
\begin{table}[h]
    \centering
    \caption{MSP430 SoC With RF Core \cite{MSP430}.}
    \label{tab:MCU}
    \begin{tabular}{ r||r|r|r|r }
        \toprule
        \textbf{Tx Power} ($P_k$) & \multicolumn{4}{c}{\textbf{Power Consumption} ($D_k$)} \\
        \midrule
        & $315$ MHz & $433$ MHz & $868$ MHz & $915$ MHz \\
        \midrule
        $14$ mW & $79.2$ mW & $100.2$ mW & $106.5$ mW & $104.4$ mW \\               
        $10$ mW & $75.6$ mW & $86.4$ mW & $99.0$ mW & $96.3$ mW \\       
        $1$ mW & $43.8$ mW & $50.4$ mW & $53.4$ mW & $52.8$ mW \\                   
        $0.25$ mW & $44.1$ mW & $52.5$ mW & $53.4$ mW & $52.8$ mW \\                 
        \bottomrule
    \end{tabular}
\end{table}

The previous table refers to an MSP430 device from Texas Instruments, a microcontroller designed with low power consumption operations and with transmission/reception capabilities. The ``Tx Power'' column represents, approximately, the power sent into the channel. The other columns describe the overall power consumption in different frequency bands. Note that, in general, $P_k \ll D_k$, regardless of the operating frequency.

\subsection{Battery Evolution}

In order to understand how the battery evolves, we briefly discuss how a frame is structured. During the first $\delta$ seconds, a stream of bits may be sent. We assume $\delta \ll T$ because the time required for a transmission is much smaller than the time required to obtain enough energy quanta for a transmission. This can be seen by comparing the values of Tables~\ref{tab:MCU} and~\ref{tab:energy_sources}. 

\vspace{0.3cm}
\begin{table}[h]
    \centering
    \caption{Energy Sources~\cite{Roundy2003,Pinuela2013,Raghunathan2005}.}
    \label{tab:energy_sources}
    \begin{tabular}{ l|r }
        \toprule
        \textbf{Energy Source} & \textbf{Power Density} \\
        \midrule
        Solar outdoors - sunny day & $15$ mW/cm$^2$ \\
        Piezoelectric & $330$ $\mu$W/cm$^3$ \\
        Vibrations & $200$ $\mu$W/cm$^3$ \\
        Solar outdoors - cloudy day & $150$ $\mu$W/cm$^2$ \\
        Ambient RadioFrequency (in London) & $6.39$ $\mu$W/cm$^2$ \\
        Acoustic Noise (at $100$ dB) & $0.96$ $\mu$W/cm$^3$ \\
        \bottomrule
    \end{tabular}
\end{table}

In Table~\ref{tab:energy_sources} we report some common energy sources with the corresponding power density levels. These values should be compared with the power consumption $D_k$ in Table~\ref{tab:MCU}. In general, except for the solar light in a sunny day, the harvested power is several orders of magnitude smaller than the transmission power. Thus, to store enough energy to perform a transmission, a long time is required. 

If a device with much smaller power consumption and a rich energy source are considered, then it may be possible to allow the device to directly send the harvested energy instead of storing it. This is particularly useful in the cases of imperfect real batteries in order to avoid the storage losses~\cite{Tutuncuoglu2015,Biason2015c} but has fewer practical applications. We leave additional study of such technique as future work.

Because of the presented situation, in our model we assume that the transmission is instantaneous and happens at the beginning of a frame (\emph{i.e.}, $\delta$ is considered negligible with respect to $T$). Under this assumption, the value $E_k$ of Equation~\eqref{eq:y_t} is replaced with $E_k-D_k$. Following the previous reasoning, the battery evolution from frame $k$ to $k+1$ can be computed as~
\begin{align}
    E_{k+1} = \min\{y_T(B_k,[E_k-D_k]^+), e_{\rm max} \}. \label{eq:batt_update}
\end{align}

Note that we explicitly take into account the effects of a finite battery. $[\cdot]^+ \triangleq \max\{0,\cdot\}$ has to be applied to $E_k-D_k$ because, in the imperfect SoC knowledge case, it may happen that a too high transmission power is selected, and $D_k > E_k$. In this case, we assume that the reward in the corresponding slot is zero (only a partial codeword is transmitted). In summary, the attained reward in slot $k$ is~
\begin{align}
    g(P_k) = \begin{cases}
        r(P_k), \quad & \mbox{if } D_k \leq E_k, \\
        0, \quad & \mbox{otherwise}.
    \end{cases}
\end{align}

\section{Optimization} \label{sec:optimization}

The system described so far can be modeled with a discrete Markov Chain (MC), whose states correspond to the actual level of charge of the battery. Since we assume that only partial SoC knowledge is available in general, the device cannot see the state of the system (partially observable MC).

Our goal is to maximize the average long-term reward $G_{\mu}$ obtained with the function $g(\cdot)$:
\begin{align}
    G_{\mu}(E_0) &= \liminf_{K \rightarrow \infty} \frac{1}{K} \mathbb{E}_{B_k,P_k}\left[\sum_{k = 0}^{K-1} g(P_k) \Bigg| E_0 \right].
\end{align}

If $g(\cdot)$ is the transmission rate, then $G$ represents the throughput of the system. We remark that in this case we are maximizing the throughput only over $\delta$ in which a device is always in idle mode for a fraction $(T-\delta)/T$ of the time. $\mu$ is the \emph{policy}, \emph{i.e.}, the function that defines which transmission power should be used. $E_0$ is the initial charge of the battery (in general we fix $E_0 = 0$).

Since we focus on the long-term optimization (that is typical for a sensor network because the device operates for long times in the same conditions), it can be shown that the problem can be formulated by exploiting the MC structure:~
\begin{align}
    G_\mu(E_0) &= \sum_{e \in \mathcal{E}} \pi_\mu(e|E_0) j_\mu(e), \label{eq:G_mu}
\end{align}

\noindent where
\begin{itemize}
    \item $\pi_\mu(e|E_0)$ is the long-term probability of being in state $e$ given the initial state of the system $E_0$;
    \item $j_\mu(e) \triangleq \mathbb{E}_{\Theta_e}[g(\Theta_e)]$. $\Theta_e$ is a random variable with probability mass function $f(\rho;e)$, $\rho \in \mathcal{P}$. Our goal is to specify the pmf of $\Theta_e$ for every $e \in \mathcal{E}$. 
\end{itemize}

When perfect SoC is available, the problem can be formulated as a Markov Decision Process (MDP) and solved with standard algorithms, \emph{e.g.}, the Value Iteration Algorithm (VIA) or the Policy Iteration Algorithm (PIA)~\cite{Bertsekas2005}. In this case, the role of $\mu$ is to define, for every possible $e \in \mathcal{E}$, the pmf $f(\cdot;e)$. In~\cite{Biason2015c} we proved that the optimal policy is deterministic, \emph{i.e.}, given the state of the system $f(\rho;e) = \chi\{\rho = \rho^\star\}$, where $\chi\{\cdot\}$ is the indicator function and $\rho^\star$ is the optimal transmission power when the state is $e$.

In the general case (imperfect SoC knowledge), the problem is a Partially Observable MDP (POMDP) \cite{Bertsekas2005}. This kind of problems can be solved using a belief state as described in Chapter~$5$ of~\cite{Bertsekas2005}. With this technique, in order to find the optimal policy, in general it is necessary to keep track of the energy arrivals in all past slots. However, this would incur a very high computational cost and, moreover, storing the optimal policy would require a large memory because all possible combinations of the past events have to be taken into account. Because of these, following~\cite{Michelusi2014}, in our work we focus on a sub-optimal approach, \emph{i.e.}, we do not consider the history and focus only on the current frame. In this context, the policy $\mu$ defines a pmf $f(\cdot;\tilde{t})$ for every partition $\tilde{t} \in \tilde{\mathcal{T}}$. Also, we restrict our study to deterministic policies as in the perfect SoC case for simplicity. Unfortunately, the problem is non-convex, thus standard optimization algorithms cannot be applied. Because of this, in our numerical evaluation, we resort to an exhaustive search (that is computationally affordable when the cardinality of $\mathcal{P}$ and the number of partitions $N$ are not too high).

\subsection{Upper Bound}

We want to find an upper bound to the performance in the case $P_k = D_k$. When $D_k > P_k$, the upper bound is still valid but looser (additional study of this case is left for future work).

Assume that $b$ energy quanta arrive. An interesting quantity to evaluate is~
\begin{align}
    a^\star &= \argmax{a} y_T(a,b),\\
    \beta^\star(b) &= y_T(a^\star,b). \label{eq:b_s_star} 
\end{align}

In practice, for a given harvesting power $b$, $\beta^\star(b)$ represents the maximum energy that can be stored in the battery (maximized over all possible values of the current state of charge, $a$). For example, using Equation~\eqref{eq:s_x_y}, we would have $a^\star < e_{\rm max}/2$. Instead, with a constant $\eta$, $\beta^\star(b) = \eta b$ and is achieved for any $a$.

Using the previous equations, it is possible to derive an upper bound for the performance.

\begin{propos}
    An upper bound to $G_\mu$ is given by~
    \begin{align}
        G_{\rm u.b.} = g(\bar{b}_s), \qquad \bar{b}_s \triangleq \sum_{b = 0}^{b_{\rm max}} p_B(b) \beta^\star(b), \label{eq:G_ub}
    \end{align}

    \noindent where $\beta^\star(b)$ is defined in~\eqref{eq:b_s_star} and $p_B(b)$ is the pmf of the energy arrivals.    
    \begin{proof}
        The energy causality constraint imposes~
        \begin{align}
            \sum_{i = 0}^k P_i \leq \sum_{i = 0}^{k-1} B_i, \qquad \forall k = 0,1,2,\ldots
        \end{align}
        
        \noindent that immediately leads to the upper bound $g(\bar{b})$ as $k$ grows to infinity~\cite{Michelusi2014}. However, in the case of storage losses, the previous inequality can be changed as follows in order to provide a better upper bound. The frame evolution is as follows~
        \begin{align*}
            &E_0 = 0 \implies P_0 = 0, \\
            &E_1 = y_T(E_0,B_0) \implies P_1 \leq y_T(E_0,B_0), \\
            &E_2 = y_T(E_0,B_0) + y_T(E_1,B_1) -P_1 \\ 
            & \implies P_1 + P_2 \leq y_T(E_0,B_0) + y_T(E_1,B_1), \\
            &\vdots \\
            &E_k = \sum_{i = 0}^{k-1} y_T(E_i,B_i) - \sum_{i = 0}^{k-1} P_i \\
            & \implies \sum_{i = 0}^k P_i \leq \sum_{i = 0}^{k-1} y_T(E_i,B_i), \quad k > 0.
        \end{align*}
        
        In order to obtain a simple upper bound, we remove the terms $E_i$ from the previous equation, assuming that they are always chosen optimally. Obviously, in the general case this is not possible and this is why, with this assumption, we only obtain an upper bound~
        \begin{align}
            \sum_{i = 1}^k P_i \leq \sum_{i = 1}^{k-1} \beta^\star(B_i), \qquad \forall k = 1,2,\ldots
        \end{align}
        
        With the previous equation, by exploiting the concavity of $g$, we obtain the upper bound~\eqref{eq:G_ub}.
    \end{proof}
\end{propos}

Note that the upper bound may depend upon the battery size. For example, focus on Equation~\eqref{eq:s_x_y}. As the battery size increases, the region where $\eta(e) \approx 1$ increases. Thus, for large batteries, $\beta^\star(b) \approx b$ and the upper bound degenerates in $g(\bar{b})$.

Instead, if $\eta$ is constant, the upper bound simply becomes $g(\eta \bar{b})$ and does not depend upon the particular battery size.

\subsection{Sub-optimal Policies}\label{sec:LCP}

In our numerical evaluation we will present results for the optimal policy. However, in general, computing the optimal policy is a difficult task, especially if the number of partitions $N$ is high (exhaustive search). Because of this, here we introduce some simpler policies that are easier to implement than the optimal one. The first, named \emph{Low Complexity Policy} (LCP), provides good performance when the battery size is not too large, whereas the second one, named \emph{Balanced Policy} (BP), works well for large batteries.

LCP is derived from the Optimal Policy (OP) with perfect SoC knowledge, namely OP$_{RP}$ (Real battery and Perfect knowledge). With perfect SoC knowledge the optimal policy can be efficiently found solving a linear program. Once the optimal RP policy is computed for every $e \in \mathcal{E}$, we define LCP as follows.

For every subset $\tilde{t} \in \tilde{\mathcal{T}}$, the power consumption of LCP in the subset $\tilde{t}$ ($d_{\rm LCP}(\tilde{t})$) is defined as the average power consumption of RP ($d_{\rm RP}(\cdot)$) over all the values $e \in \tilde{t}$, \emph{i.e.}, $d_{\rm LCP}(\tilde{t}) = 1/|\tilde{t}| \sum_{e \in \tilde{t}} d_{\rm RP}(e)$.\footnote{An alternative definition of LCP would consider a \emph{weighted} mean (\emph{e.g.}, with the steady-state probabilities). However, we have numerically verified that this leads to worse performance.} LCP can be seen as the adaptation of OP$_{RP}$ to the imperfect knowledge case. We will show that this approximation is legitimate if the battery is small, but degenerates quickly as the battery size increases. This remarks the importance of using ad hoc policies for the cases of imperfect SoC knowledge with real batteries.

The aim of BP is to consume, on average, a power equal to $\bar{b}_s$ (Equation~\eqref{eq:G_ub}). If this were always possible, then BP would achieve the upper bound. The definition of BP strictly depends upon the number of partitions $N$ and the battery inefficiencies (function $\eta(e)$). In the case $N = 2$ (LOW-HIGH) and for $\eta(e)$ defined in Equation~\eqref{eq:s_x_y} we define BP as $d({\rm LOW}) = 0$ and $d({\rm HIGH}) = \bar{b}_s$. Note that it is important to set $d({\rm LOW}) = 0$ in order to avoid to be trapped in the low energy states (we will discuss this effect in more detail in the next section).

\section{Numerical Results} \label{sec:numerical_evaluation}

In our numerical results we discuss the behavior and compare the performance of the policies in Table~\ref{tab:policies}.

\begin{table}[h]
    \centering
    \caption{Policies.}
    \label{tab:policies}
    \begin{tabular}{ c|c|c }
        &\textbf{Perfect SoC} & \textbf{Imperfect SoC} \\
        \hline
        \textbf{Ideal Battery} & OP$_{\rm IP}$ & OP$_{\rm II}$ \\
        \hline
        \textbf{Real Battery} & OP$_{\rm RP}$ & OP$_{\rm RI}$, BP, LCP, OP$_{\rm II}$ \\
    \end{tabular}
\end{table}

Where OP$_{\rm i}$ represents the optimal policy in scenario i. Our focus is on the RI case.\footnote{OP$_{\rm IP}$ was studied in~\cite{Michelusi2012}, OP$_{\rm II}$ in~\cite{Michelusi2014} and OP$_{\rm RP}$ in~\cite{Biason2015c}.} Here, we also evaluate the performance of OP$_{\rm II}$ applied to this scenario and of BP and LCP, defined in Section~\ref{sec:LCP}.

\subsection{Optimal Policies Structure}

We now discuss the shape and the performance of the four optimal policies of Table~\ref{tab:policies}. A channel gain $\Lambda = 10^{-2}$ is chosen in order to operate in the low SNR regime~\cite{Sharma2010}. Also we set $\tilde{\mathcal{T}} = \{{\rm LOW},{\rm HIGH}\}$ ($N = 2$) and neglect the circuitry costs, thus the power consumption in state $e$ is equal to $\rho(e)$. The storage losses are modeled with Equation~\eqref{eq:s_x_y}. The battery size is set to $e_{\rm max} = 100$ energy quanta and the energy arrival process is described by a truncated geometric random variable with mean $\bar{b} = 20$ and maximum arrivals $b_{\rm max} = 50$.

\begin{figure}[t]
  \centering
  \includegraphics[trim = 0mm 0mm 0mm 3mm,  clip, width=1\columnwidth]{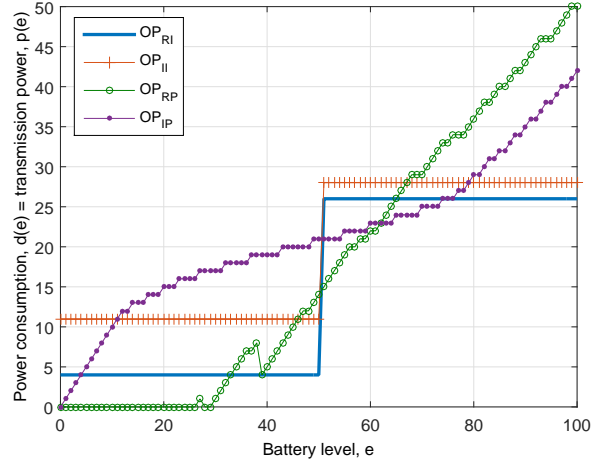}
  \caption{Power consumption $d(e)$ (without circuitry costs $d(e) = \rho(e)$) of the optimal policies in Table~\ref{tab:policies} as a function of the battery level $e \in \mathcal{E}$.}
  \label{fig:policies}
\end{figure}

In \figurename~\ref{fig:policies} we plot the transmission power $\rho(e)$ as a function of the battery size $e$. Note that, because of the battery imperfections, the curve of OP$_{\rm RP}$ (Real battery with Perfect knowledge) is not as smooth as the one for the ideal case OP$_{\rm IP}$ (Ideal battery with Perfect knowledge). 

Interestingly, for the low energy states, the curves for a real battery are lower than their counterparts for an ideal battery. In practice, when the battery is almost empty, the energy losses are high, thus only a small fraction of the harvested energy can be stored. In this case, it is better not to transmit (or transmit with low power) and wait until the battery reaches a more favorable region. 

We now want to highlight the importance of applying the right optimal policy. In particular, we show that applying OP$_{\rm II}$ (Ideal battery with Imperfect knowledge) and OP$_{\rm RP}$ (Real battery with Perfect knowledge) is sub-optimal in the RI case.

\subsubsection{OP$_{\rm II}$ applied to RI}
It can be verified that applying OP$_{\rm II}$ to the RI case provides a reward equal to zero whereas OP$_{\rm RI}$ provides a reward of $0.1655$ (we discuss how the reward changes in Section~\ref{sec:num_eval_throughput}). This result can be explained as follows. In state $e = 0$, the maximum storable energy is given by $y_T(0,b_{\rm max})$. In our example, $b_{\rm max} = 50$ and, using Equation~\eqref{eq:s_x_y}, it can be verified that $y_T(0,b_{\rm max}) = 6.3$, \emph{i.e.}, approximately $6$ energy quanta. Starting from $E_0 = 0$, after the first slot we have $E_1 = 6$. In this slot, a transmission power corresponding to $11$ energy quanta is demanded (as shown by the red curve), but, since $E_1 < 11$, only $6$ energy quanta are drawn from the battery. However, this results in a failed transmission (the transmission is interrupted at $55\%$), thus the corresponding reward is zero. In the successive slot, $E_2 = 0$ and the process repeats periodically, providing a global reward equal to zero. Note that this behavior is not influenced by the initial state: even if the battery were initially fully charged, there would exist a positive probability of reaching state $0$ and being trapped there. A similar behavior can be noticed in the perfect SoC case, but with less disruptive effects. This example highlights an important characteristics of real batteries, that has to be accurately taken into account when a system is designed.

\subsubsection{OP$_{\rm RP}$ applied to RI} \label{sec:RP_to_RI}
The optimal policy with perfect knowledge cannot be directly applied to the imperfect SOC case. Therefore we have to resort to the approximation described in Section~\ref{sec:LCP}, namely LCP. OP$_{\rm RP}$ takes into account the fact that the low energy states should be avoided in order not to trap the battery level. However, by taking the average over every subset, it may happen that, as in the OP$_{\rm II}$ case, a too high power consumption is employed, resulting also in this case in zero reward. This behavior is better explained in Section~\ref{sec:num_eval_throughput}.

\begin{figure}[t]
  \centering
  \includegraphics[trim = 0mm 0mm 0mm 3mm,  clip, width=1\columnwidth]{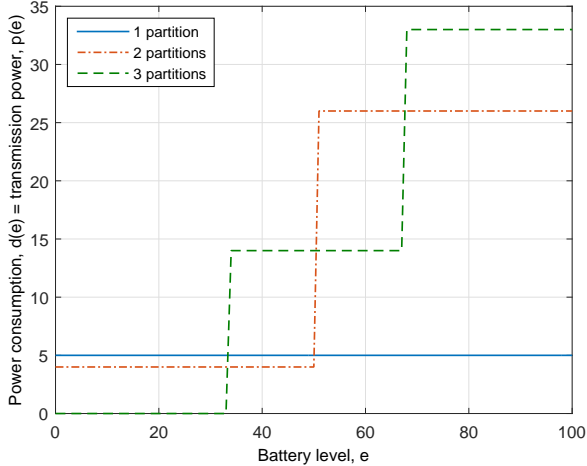}
  \caption{Power consumption $d(e)$ (without circuitry costs $d(e) = \rho(e)$) of OP$_{\rm RI}$ for $N = 1,2,3$ as a function of the battery level $e \in \mathcal{E}$.}
  \label{fig:plot_Real_Imperfect}
\end{figure}

While applying a sub-optimal policy to the RI case may result in zero reward, OP$_{\rm RI}$ takes into account this effect and avoids the inefficient energy states. This is easy to obtain when $N$ is high. However, if $N$ is very low ($N = 1$, \emph{i.e.}, no SoC knowledge) this may result in a significant performance degradation. In \figurename~\ref{fig:plot_Real_Imperfect} we plot the transmission power of OP$_{\rm RI}$ when we use one, two or three partitions. Note that in the case $N = 1$, in order to avoid the effect previously described (reward equal to zero if the power consumption is too high), $\rho(e)$ has to be very low. With $N = 3$, any transmission is avoided for the low energy states in order not to operate in an inefficient battery region. The rewards obtained in the three cases are $0.0488$, $0.1655$ and $0.1670$, respectively. Also, the reward obtained with perfect SoC knowledge is $0.1714$. We notice that even a rough idea of the battery SoC (quantization to two levels, LOW and HIGH) provides close-to-optimal performance, and therefore a more accurate representation is not necessary. On the other hand, if no information is provided ($N=1$) the performance is heavily sub-optimal.

\begin{figure}[t]
  \centering
  \includegraphics[trim = 0mm 0mm 0mm 3mm,  clip, width=1\columnwidth]{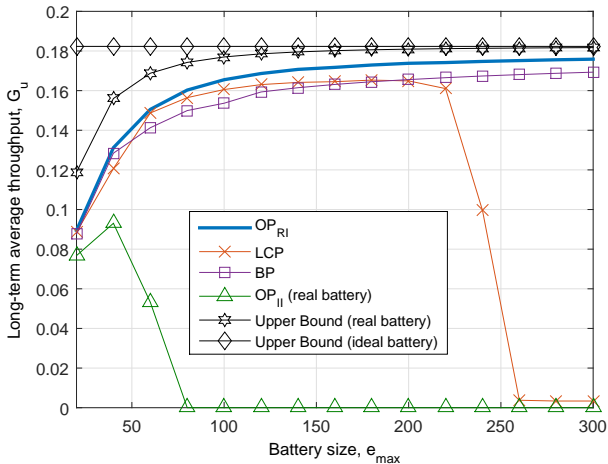}
  \caption{Long-term average throughput $G_\mu$ (Equation~\eqref{eq:G_mu}) as a function of the battery size $e_{\rm max}$ (with $\bar{b} = 20$).}
  \label{fig:plot_e_max}
\end{figure}

\subsection{Throughput} \label{sec:num_eval_throughput}
Another interesting quantity to analyze is the throughput of the various policies when the battery size increases. In this case, see \figurename~\ref{fig:plot_e_max}, we also want to show the system upper bounds and the performance of LCP and BP. As expected, as the battery size increases, the optimal policy and the upper bound~\eqref{eq:G_ub} converge to $g(\bar{b})$ (upper bound with an ideal battery). An interesting point is that $G_{\rm u.b.}$ is much closer to the performance of the optimal policy than $g(\bar{b})$, especially for low batteries. 

We remark that $G_{\rm u.b.}$ is an upper bound also for the RP (Real battery with Perfect knowledge) case. Thus, since the upper bound and the performance of OP$_{\rm RI}$ are quite close, we can state that it is almost equal to considering $\tilde{\mathcal{T}}$ with a LOW-HIGH subsets, or the perfect knowledge case ($|\tilde{\mathcal{T}}| = e_{\rm max}+1$), \emph{i.e.}, even considering $\tilde{\mathcal{T}}$ with very few subsets it is still possible to achieve close to optimal rewards.

The figure also reports the Balanced Policy. Its performance is very close to the reward of OP$_{\rm RI}$, for both low and high batteries. BP also converges to the upper bound $G_{\rm u.b.}$ for very large batteries. This is because, on average, BP transmits data with power $\bar{b}_s$.

Finally, we applied the optimal II (Ideal battery with Imperfect knowledge) and RP (Real battery with Perfect knowledge) policies to the RI scenario. As pointed out in Section~\ref{sec:RP_to_RI}, in order to apply RP to the RI case, we introduced LCP. As can be seen from the figure, OP$_{\rm II}$ achieves a very low reward that quickly degenerates to zero. Instead, LCP seems to work well for low batteries but as soon as the battery size grows too much, its reward degenerates to zero as well. This further emphasizes the importance of using the proper policy when a real battery with imperfect knowledge is considered.

\subsection{Real Case Example}

\begin{table}[H]
    \centering
    \caption{Numerical example parameters.}
    \label{tab:parameters}
    \begin{tabular}{ l|r }
        \toprule
        \textbf{Parameter} & \textbf{Value} \\
        \midrule
        One energy quantum & $10\ \mu$J \\
        \hline
        Frame length ($T$) & $1$ s \\
        \hline
        Slot length ($\delta$) & $5$ ms \\
        \hline
        Available bandwidth ($W$) & $2$ MHz \\
        \hline
        Noise power density ($N$) & $10^{-20.4}$ W/Hz \\
        \hline
        Channel state ($H$) & $3\cdot 10^{-13}$ W/Hz \\
        \hline
        \multirow{2}{*}{Min/Max energy consumption ($d(e)$)} & $44.1/106.5$ mW \\
        & ($22/53$ quanta) \\
        \hline
        \multirow{2}{*}{Min/Mean/Max harvested energy ($b(e)$)} & $10/300/500\ \mu$J \\
        & ($1/30/50$ quanta) \\
        \bottomrule
    \end{tabular}
\end{table}

We present numerical results for the device reported in Table~\ref{tab:MCU}. In Table~\ref{tab:parameters} we summarize the parameters used for our example.

The considered energy source can be, \emph{e.g.}, piezoelectric or vibrations. In order to model the statistics of the energy source we use a truncated Poisson random variable in order to have a distribution centered around its mean. Note that $T \gg \delta$. The instantaneous reward function is defined according to Shannon's formula~
\begin{align}
    g(\rho) = \frac{\delta}{T} W \log_2\left(1+\frac{H}{W N_0} \rho\right).
\end{align}

\begin{figure}[t]
  \centering
  \includegraphics[trim = 0mm 0mm 0mm 3mm,  clip, width=1\columnwidth]{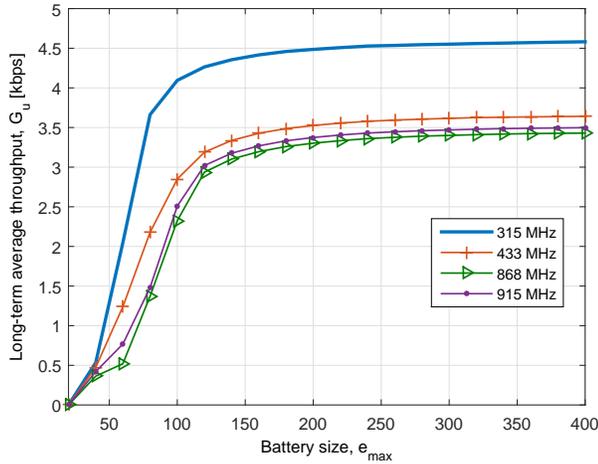}
  \caption{Long-term average throughput $G_\mu$ (Equation~\eqref{eq:G_mu}) as a function of the battery size $e_{\rm max}$ (with $\bar{b} = 30$) for the various frequency bands of Table~\ref{tab:MCU}. The bandwidth is $W = 2$ MHz.}
  \label{fig:plot_e_max_real}
\end{figure}
The term $\delta/T$ takes into account the fact that the optimization is performed in the first $\delta$ seconds of the frame.

We assume that the device fixes a priori the transmission frequency according to the columns of Table~\ref{tab:MCU}. In \figurename~\ref{fig:plot_e_max_real} we show the throughput for a $2$ MHz bandwidth, expressed in kbps, at the four frequencies. At $315$ MHz, the throughput is greater because, for every fixed transmission power $\rho$, the corresponding power consumption $d$ is lower.

\section{Conclusions} \label{sec:conclusions}

In this paper we studied the throughput optimization problem of an Energy Harvesting Device with a Markov Decision Process approach. We explicitly considered the effects of imperfect batteries and in particular the imperfect SoC knowledge and the energy storage inefficiencies (with losses related to the SoC status). We based our model on some realistic consideration about the transmission/consumption/harvesting power. We found a performance upper bound and showed that, with a real battery, the upper bound and the performance are lower than in the ideal case. We proposed low complexity policies and showed that BP is a good approximation for the optimal case. We discussed the application of the policies derived in traditional settings and showed that they are strongly sub-optimal and that, in some cases, the corresponding throughput could even be zero. This emphasizes the importance of considering the real characteristics of an EHD when the optimization process is performed.

As part of the future work we would like to set up a model where also an ultra-low power device can be studied, \emph{i.e.}, $\delta$ and $T$ are similar. Also, other model imperfections can be included, \emph{e.g.}, energy leakage or battery degradation.

\bibliography{../../../EHD}{}
\bibliographystyle{IEEEtran}

\end{document}